\documentclass[12pt]{article}

\usepackage{graphicx,amssymb,multirow}

\def\be{\begin{equation}}
\def\ee{\end{equation}}
\def\ba{\begin{eqnarray}}
\def\ea{\end{eqnarray}}

\def\nn{\nonumber}

\begin{document}

\title{A classical attack on the coherent one way protocol for quantum key distribution}
\author{Michail Stoilov\\ 
{\small Institute for Nuclear Research and Nuclear Energy, }\\
{\small Bulgarian Academy of Sciences,} \\
{\small Sofia 1784, Tsarigradsko shosse 72, Bulgaria }\\
{\small email: mstoilov@inrne.bas.bg}}
\maketitle
\begin{abstract}
We propose a way to retrieve the secure key generated by the coherent one way protocol without reading the information transmitted on the quantum channel. 
\end{abstract}

\section{Introduction}
The coherent one way (COW) protocol \cite{cow0} is specifically designed for quantum key distribution (QKD) \cite{qkd} between two distant partners, traditionally called Alice and Bob.
It is not suitable for data transfer and this is intentional.
Bob in principle can receive only very few of the bits sent by Alice
 for the key construction.
Moreover, the received bits are randomly chosen from the transmitted ones.
Therefore, if an eavesdropper, say Eve, wants to know the entire key generated by Alice and Bob,  she has to be able to detect all bits transmitted by Alice. 
In other words, there has to be data transfer from Alice to Eve, something which is specifically designed not to be possible under COW protocol.

The protocol is tested against several types of eavesdropping attacks \cite{cow0, cow_err1}  most of them being variants of the photon number splitting (PNS) attack \cite{pns1, pns2}.
Here our approach to attack COW protocol is different.
Eve does not detect anything.
She only restricts (in a controlled way) the information received by Bob.
Finally, she is able to retrieve  the generated key from the public information exchange between Alice and Bob.

\section{COW protocol}
According to COW protocol \cite{cow0} the partners are connected by two channels: a ``classical channel''  and a ``quantum channel''.
The difference between both channels is that it is possible to  read the information on the classical one  without distorting it, while the reading of  information on the quantum channel alters it.

The classical channel is two directional.
It is used prior the transmission on the quantum channel in order to synchronize Alice's and Bob's clocks.
After the transmission (of the information via the quantum channel) the classical channel is used  by Bob to reveal the  successive number of the bits he have received, and by Alice to pick which bits from the received ones to be made public for verification and which to be used in the key.
Alice announces as well some special time moments to be checked for coherence.

The quantum channel is one directional, from Alice to Bob.
Alice generates random bit sequence (raw key information, or just information) and uses pulse modulation to transmit it.
In other words, any information bit (IB) is encoded  as  two pulses, or transmission bits (TBs), each with fixed duration $\tau$.
The TB $0_{trans}$ is represented by a time bin where no photons are transmitted and $1_{trans}$ is represented by a time bin with $\mu$ mean photon number and with Poisson photon number distribution in it. 
In other words
\ba 
0_{trans}&=&| 0\rangle\nn\\
1_{trans}&=& | \mu \rangle = e^{-\mu/2} \sum_{n=0} \frac{\mu^{n/2}}{\sqrt{n!}}\,| n\rangle, \label{dmu}
\ea
where $| n\rangle$ is the normalized state with $n$ photons. 
The encoding used by COW protocol is:
\ba 
0_{info} &\rightarrow& 10_{trans}\nn\\
1_{info} &\rightarrow& 01_{trans},\label{enc}
\ea
where the left transmission bit is transmitted first.
The encoding (\ref{enc}) ensures that Bob detects with equal probability $0_{info}$ and $1_{info}$ provided his detector is noiseless or without dead time.

According to COW protocol Bob determines the value of a received IB by the time he detects a photon. 
For instance, if the transmission starts at $t=0$ and if Bob detects a photon in the time interval $T\in(2(n-1)\tau,(2n-1)\tau)$ he knows that this is the $n$-th successive  IB and its value is $0_{info}$.

Occasionally, Alice transmits  the so called decoy sequence $11_{trans}$. 
This sequence (which presents in $10_{info}$ as well) is used to detect PNS attack.
For this purpose Bob randomly selects a fraction $(1-t_B) \ll 1$ of the  incoming photons which  are directed not to Bob' data detector but to an unbalanced interferometer tuned for destructive interference of $11_{trans}$.
The interference picture is monitored by a detector and if the key generation is subjected to PNS attack which destroys the $11_{trans}$ coherence this detector will fire.
Note, that most of the decoy sequences (about $t_B(2-t_B)$ of them) do not pass through the  interferometer, but through Bob's data detector and those which are detected are misinterpreted by Bob as $0_{info}$ or $1_{info}$.
The reason is that Bob can not detect two consecutive TBs because of detector dead time.
These errors are corrected during the Alice and Bob dialog on the classical channel.

It follows from Eqs.(\ref{dmu}) that $| 0\rangle$ and $| \mu \rangle$ are not orthogonal
\be \langle 0|\mu\rangle=e^{-\mu/2}\ne 0,\label{noort}\ee
and therefore, Bob, and Eve as well, can misinterpret $1_{trans}$ as $0_{trans}$.
This fact puts limitation on the usage of COW protocol for data exchange, 
because in the protocol $\mu = 0.5$.
\begin{figure}[h]
\begin{center}
\includegraphics[scale=.72]{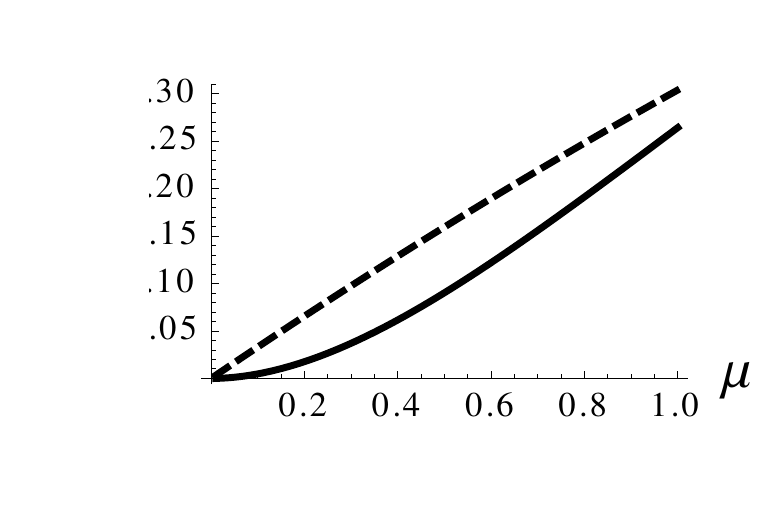}
\caption{\small Solid line: The fraction of the transmitted $1_{trans}$ which can be detected with an ideal single photon detector (i.e. they are with photon number $\ge 1$) as a function of mean photon number $\mu$ in the pulse. Dashed line: the fraction of the detectable $1_{trans}$ which can be split (i.e. with photon number $\ge 2$).
}\label{mu_choice}
\end{center}
\end{figure}
As it is seen from Fig.\ref{mu_choice} this is not the only possible choice with smaller $\mu$ trading speed for security.
In fact smaller  $\mu$ is indeed used --- because of the quantum channel transmission coefficient $t<1$ (for large distances between Alice and Bob $t \ll 1$) and the facts that Bob send only a fraction $t_B<1$ of incoming photons to his data detector which is with  quantum efficiency $\eta\ll 1$  the effective mean photon number at Bob's side is much smaller than the initial one at Alice's side
\be \mu_{eff}=\mu\,t\,t_B\,\eta \ll \mu. \ee
Note that the minimal $\mu_{eff}$ for which the protocol can work depends entirely on Bob's data detector noise.
Surely, smaller $\mu_{eff}$ means lower key generation rate. 
If we take into account the detector dead time this rate will be reduced further, in some cases --- by several orders of magnitude.

In the practical  setups \cite{cow1, cow2} Alice prepares the transmission sequence using adjustable continuous wave laser and intensity modulator.
Alice monitors the laser power continuously to ensure $\mu=0.5$.
The pulse duration is $\tau\approx 2.4$ns.
The modulator transmits the laser light during $1_{trans}$ and  blocks it completely during $0_{trans}$.
The fraction of incoming photons directed to Bob's data detector is $t_B=0.9$.
There are two reported algorithms for generation of the decoy sequence:
(a) the  decoy sequence substitutes substring $1010_{info}$  whenever it appears in the raw key information \cite{cow2}, or (b) there is a decoy sequence at the start of every $8$ TBs  \cite{cow1}.
Fortunately, as we shall see in the next section, the exact mechanism for decoy state generation is not important for the proposed attack.

\section{The attack}

The proposed attack is based on the COW feature that Bob receives only small fraction of the information transmitted by Alice. 
We suppose that Eve has full access to the information on the classical channel.
So, her clock is synchronized with Alice's and Bob's ones and she is aware which are the successive number of the IBs received by Bob and which of them have been used for the key.

\begin{table}[h]
\begin{center}

\begin{tabular}{|c|c|c|c|c|c|}
  \hline
 \multicolumn{2}{|c|}{Alice}&\multicolumn{2}{c|}{Eve}&\multicolumn{2}{c|}{Bob}\\
\hline
IB& TBs& IB & TBs& received bits & IB\\
\hline
0&10& 0&10& 10&0\\
0&10& 1&01& 00&n.d.\\
1&01& 0&10& 00&n.d.\\
1&01& 1&01& 01&1\\
d.s.&11& 0&10& 10&0\\
d.s.&11& 1&01& 01&1\\
 \hline
\end{tabular}
\caption{\small The result of Eve's attack is in Bob's columns. Here, ``d.s.'' stands for decoy state, and ``n.d.'' is for not defined state.}
\label{attack}

\end{center}
\end{table}

The attack on the quantum channel is performed synchronously with the information transmission from Alice to Bob.
Eve does not try to read this information, because this will be detected, but altered it.
For this purpose she generates her own random\footnote{In fact the sequence may be just $101010\dots$.} bit sequence, translates it into transmission sequence, and synchronously applies bitwise AND operation between this sequence and the signal on the quantum channel.
In the practical  setups this can be done by an intensity modulator, analogous to the one used by Alice.
As a result (see Table \ref{attack}) with $50 \%$ probability any IB  generated by Alice passes trough Eve unchanged, and  with $50 \%$ probability it becomes the sequence $00_{trans}$ which is not defined but is highly expected in COW protocol.
All decoy sequences are destroyed and substituted by  $01_{trans}$ or $10_{trans}$ sequences.
As it has been explained above, Bob always misinterprets the detected decoy sequences, so their substitution with  $0_{info}$ or $1_{info}$ can not be noticed.

The crucial moment of the attack is that Bob can detect only these IBs generated by Alice which coincide with the generated by Eve ones.

At the end of the transmission Bob announce on the classical channel the sequential number of the IBs he has received.
This is the moment in which Eve effectively detects the information received by Bob, because she knows the time sequence of her own bits.
The next steps performed by Alice and Bob are completely classical and public, so at the final Eve will have a copy of their key.

\section{Conclusions}
We propose a  completely classical attack on COW protocol for QKD in which Eve does not try to detect anything on the quantum channel. 
She only superpose her own information upon the transmitted one.
Eve is able to retrieve the shared key only after the end of quantum communication, on the base of the data  sent between Alice and Bob via the classical channel.

\end{document}